\documentclass[amssymb,twocolumn,prb,showpacs,aps]{revtex4}
\usepackage{bm}
\usepackage{graphicx}

\begin{document}

\title{ Correlations of  spin-polarized and entangled electrons with Berry phase}
\author{Xuean Zhao$^{1,2}$,  Hui Zhao$^{1}$, Pei Wang$^{1}$, and You-Quan Li$^{1}$}
\address{$^{1}$Zhejiang Institute of Modern Physics, Zhejiang University, Hangzhou 310027, China\\
$^{2}$National Laboratory of Solid State Microstructures, Nanjing
University, Nanjing 210093, China}

\begin{abstract}
The correlation and fluctuation of both entangled electrons and
spin-polarized pairs affected by two rotating magnetic fields in a
setup proposed by J. Carlos Egues etc. (Phys. Rev. Lett. {\bf
89}(2002) 176401) are studied theoretically by using scattering
approach. Differing from polarized pair, the entangled electron
pairs are shown to behave like a composite particle with the total
spins and its $z$ components. The singlet and  triplet states
exhibit different bunching and antibunching features, which can be
easily adjusted by the magnetic fields. The correlations and
variances can show up distinguish output signals for the four
incident states. Our results are expected to be tested by using
coincident technique.

\end{abstract}


\maketitle

\section{introduction}

Entanglement and non-locality of EPR (Einstein-Podolsky-Rosen)
pairs \cite{epr} or Schr{\"o}dinger cat states\cite{schrodinger}
are the most intriguing and profound features of quantum
mechanics. They give rise to, even in the absence of interaction,
correlations between spatially separated  particles, which can not
be described as a product of the quantum states of the two
particles, and can not be explained by any local hidden variable
and realistic theory.\cite{bell}  However, these nonclassical and
non-local correlations of entanglement have been tested
convincingly by Bell-inequality-type measurements in
optics.\cite{zeilinger,bel} Besides the fundamental aspects, a
great deal of efforts have been made to utilize the properties of
entangled particles for quantum information, quantum computation
or other potential applications~\cite{bennett}. It is foreseeable
that the realistic and applicable quantum devices should be
composed of solid state elements.  Therefore, it is significant
and desirable to understand the entangled charged particles.

Recently, much interest has been taken in entanglement of
electrons in solid state
environment.\cite{buttiker1,buttiker2,buttiker3,loss1,loss2,loss3,loss4,lesovik1,lesovik2}
Unlike photons, electrons carry charge and spin. The Coulomb
interaction between electron and external electric field  makes it
flexible to be manipulated. Moreover, recent experiments showed
that electron spin in a semiconductor has a relative long
dephasing time, approximately microseconds, and it can be
transported coherently over 100 micrometers\cite{kik}. These
properties make electron a promising candidate for quantum
applications. A few proposals based on electrons showed how to
create an entangled pair of electrons in solid state systems, {\it
e.g.}, superconductors\cite{recher,lesovik3} or quantum
dots\cite{loss3,loss4}, as a source of entangled beams of
electrons.

To detect and manipulate the spin entanglement, the electric
current noise of a beam splitter as a test of spin entanglement
was proposed by Samuelsson {\it et al}\cite{buttiker1}, Burkard
{\it et al}\cite{loss4} and Carlos {\it et al}. In these schemes,
the orbital symmetry of the entangled electron wave function can
be probed directly by the current noise (shot noise). Since the
total wave function consists of orbital part and spin part, the
symmetry of the orbital degree of freedom is intrinsically related
to that of the spin via the Pauli exclusive principle or the
parity requirement for the wave function for indistinguishable
particles~\cite{liu}. This intrinsic relation between orbital and
spin in the two-electron wave function imposes an essential
connection between the spin and the orbital for the entangled
electrons. If one alters the spin symmetry of the pair wave
function, it will definitely change the orbital symmetry of the
total wave function\cite{loss5}.

Geometric phase such as Berry phase plays an important role in
quantum systems with cyclic evolution. Particularly, in nanoscale
electronic devices~\cite{zhou}, a surprising effect is that a
quantum system retains a memory of its motion when it undergoes a
cyclic evolution~\cite{berry1,sjo1}. In quantum information
processing, Berry phase shows purely geometric characteristics
that can be used to tolerate external parameter
fluctuations\cite{flun1,nmr1}. It has been shown that Berry phase
can be used to implement conditional phase shifts, which gives
quantum computing an ability to execute conditional dynamics
between two quantum bits (qubit), where the state of one qubit
affects the evolution of the other qubit during a quantum
computation~\cite{eprin,zhu}. Actually, Berry phase has been
demonstrated in many systems, such as NMR~\cite{nmr1,nmr2,nmr3},
neutron beams~\cite{neutron1,neutron2} and
nanostructures\cite{nano1,nano2}, which are merely in single
particle literature. Sj\"oqvist considered the geometric phase for
entangled spin-$\frac{1}{2}$ pairs in a time-independent uniform
magnetic field\cite{sjo3}, and Tong {\it el al} considered the
geometric phase for entangled states of two spin-$\frac{1}{2}$
particles in rotating magnetic field~\cite{tong}. They showed that
the geometric phase acquired by one of the entangled particles is
always affected by the other particle which is even a free state.
However, for unentangled particles, the geometric phase for the
product state equals the sum of geometric phases acquired by each
particle\cite{tong1}. Thus studies on entangled quantum systems
are still in its infancy.

   In this paper we study the influence of the
   Berry phase on the entangled states and polarized states
   of two spin-1/2 electrons by taking account of
   the correlation of the two electrons.
   We adopt the same approach as in Ref.\cite{loss5} and take for granted the existence
   of an entangler constituted by a superconductor and quantum dots \cite{loss3,recher}
   that are in Coulomb blockade regime.
   The setup shown in Fig.1 consists of an entangler and a beam splitter.
   In this scheme, two electrons incoming from lead 1 and 2 are scattered to
   lead 3 and  4.
   Berry phase is generated by imposing an adiabatically rotating magnetic
   field on one of the paths of the incoming leads.
   The geometric phase is associated with nonvanishing dynamical phase generally.
   The dynamical phase ought to be diminished by some means for that  it will spoil the effect
   of Berry phase.
   We use the well-developed technique,
   spin-echo approach~\cite{nmr1,echo1,echo2,echo3},
   to compensate the dynamical phase and make the states sensitive to Berry phase.

The EPR pairs of electrons emitted from the entangler are
generally singlet or triplet states~\cite{buttiker1,loss3,loss5}.
In the scheme of Ref. \cite{loss4} three triplet states can not be
distinguished from each other due to the same symmetries of
spatial wave functions {\it i.e.}, the same anti-bunching
behaviors. The improved approach~\cite{loss5} can distinguish the
entangled and spin polarized beams by imposing a Rashba
interaction on one of the pair electrons. For the spin polarized
electrons, only the polarization along $z$-axis is nonzero in the
current autocorrelation (shot noise). The electrons polarized
along $y$-axis show noiseless. It also shows that  $y$ component
of an entangled triplet $\left |T_{e_{y}}\right \rangle$ and $z$
component of unentangled triplet $\left |T_{u_{z}}\right \rangle$
are noisy while the triplets of entangled $z$ component $\left
|T_{e_{z}}\right \rangle$ and unentangled $y$ component $\left
|T_{u_{y}}\right \rangle$ are noiseless. We propose in this paper
an approach to demonstrate the correlation of two electrons in
both singlet and triplet states utilizing rotating magnetic
fields. In our scheme, we employ two reversed rotating magnetic
fields for one of the two electrons passing through so that their
dynamic phase can be cancelled. We indicate that the correlation
of the two electrons shows different values for entangled singlet
state, triplet state as well as unentangled polarized states. In
addition, the values of correlations can be changed by Berry
phase, which gives bunching and antibunching as well as
intermediate behavior as a function of rotating magnetic field
parameters.

\section{Model and  formulation}

The Hamiltonian of an electron in  rotating magnetic fields is
\begin{equation}
H=H_{0}-\frac{1}{2}g\mu_{B} \bm{\sigma }\cdot
\mathbf{B}(t),\label{eq1}
\end{equation}
where $H_0$ represents a free motion and the second term
represents the Zeeman energy; $\mu_{B}$ is the Bohr magneton, $g$
the Land\'{e} factor and $\bm\sigma$ the Pauli matrices. Here
${\bf {B}}(t,x)=\mathbf{B}(t)\delta_{xx_1}-\mathbf
{B}(t)\delta_{xx_2}$ are time-dependent magnetic fields located at
$x_1$ and $x_2$, respectively, and $\delta_{xx_i}$ denotes the
Kronecker delta function. The orientations of the magnetic fields
are opposite in direction $\vec{n}(\theta)$, as shown in Fig.1.
The rotating magnetic fields can be formed by two orthogonal
magnetic fields, {\it i.e.}, one is along $z$ axis with fixed
magnitude while the other is rotating in $x$-$y$ plane. An
electron passing through such magnetic fields acquires both
dynamical and geometric phases. When the magnetic field rotates
slowly or adiabatically, the time evolution of the state obeys
\begin{figure}[tbph]
\vbox to2.2in{\rule{0pt}{2in}} \includegraphics{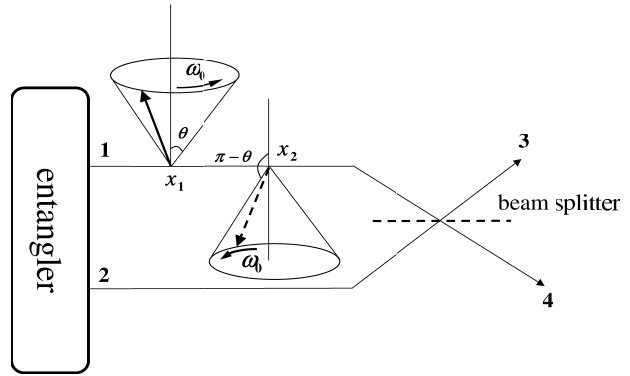}
\caption{A  model experiment setup. The entangler emits two
electrons, which are singlet or triplets. Two reversed rotating
magnetic fields are located at $x_1$ and $x_2$. The dashed line
represents a beam splitter. The incoming channel 1 and channel 2
are separated in space, the exit channels are 3 and 4.
}\label{fig1}
\end{figure}
\begin{equation}
\left| \Uparrow_{n} ;t\right\rangle =\exp \left\{ -\frac{i}{\hslash }\int E_{n}(%
\mathbf{B}(t^{\prime }))dt^{\prime }\right\} \exp (i\gamma
_{n}(t))\left| \Uparrow_{n}(t)\right\rangle,
\end{equation}
where $n$ is the direction of total magnetic field in orientation
$\vec{n}(\theta;t)$. $E_{n}=E_{0}+E_{\pm }$ is an instantaneous
eigenenergy in the direction of $\vec{n}(\theta;t)$. Here
$E_{\pm}=\mp \frac{1}{2}g\mu_{B} B$ and the geometric phase is
\begin{eqnarray}
\nonumber \gamma _{n}(t) &=&i\int_{\mathbf{B}(0)C}^{\mathbf{B}(t)}
\left\langle \Uparrow_{n} (\mathbf{B}(t^{\prime }))\right| \nabla
_{\mathbf{B}}\left| \Uparrow_{n} (\mathbf{B}(t^{\prime
}))\right\rangle
d\mathbf{B}(t^{\prime }) \\
&=&i\int_{0C}^{t}\left\langle \Uparrow_{n} (t^{\prime })\right| \frac{\partial }{%
\partial t^{\prime }}\left| \Uparrow_{n} (t^{\prime })\right\rangle dt^{\prime
},
\end{eqnarray}
where $C$ is the contour of a path in parameter space. The
instantaneous eigenstates of the Hamiltonian (\ref{eq1}) at time
$t$ can be expanded in the $\sigma_z$ basis as
\begin{eqnarray}
\label{zeign1} \left| \Uparrow_{n} (t)\right\rangle &=&\cos
\frac{\theta }{2}\left| \uparrow \right\rangle +\sin \frac{\theta
}{2}\text{e}^{i\omega _{0}t}\left|
\downarrow \right\rangle \\
\left| \Downarrow_{n} (t)\right\rangle &=&-\sin \frac{\theta
}{2}\left| \uparrow \right\rangle +\cos \frac{\theta
}{2}\text{e}^{i\omega _{0}t}\left| \downarrow \right\rangle,
\label{zeign2}
\end{eqnarray}
where $\left| \uparrow \right\rangle $ ($\left| \downarrow
\right\rangle $) is eigenstate for spin up (spin down) in $z$
axis. $\theta$ is the angle between the directions of the total
magnetic field and the $z$ axis. The level splitting in the field
is
\begin{equation}
E_{\pm}=\mp \frac{1}{2}g\mu B=\mp \hslash \omega _{1}, \quad
E_{-}-E_{+}=2\hslash \omega _{1}.
\end{equation}

The adiabatic condition requires
\begin{equation}
\frac{\omega _{0}}{\omega _{1}}\ll 1,
\end{equation}
where $\omega_{0}$ is the frequency of the rotating magnetic
fields and $\omega_{1}$ the frequency of Rabi oscillation between
the two levels. After a period $\tau =2\pi/\omega _{0}$ the
eigenstate $\left| \Uparrow_{n} ;t=0 \right\rangle$($\left|
\Downarrow_{n} ;t=0 \right\rangle$) subjects to one  rotating
magnetic field and evolves to
\begin{eqnarray}
\nonumber \left| \Uparrow_{n} ;t=\tau \right\rangle
=\text{e}^{i\gamma _{+}(\theta )}\text{e}^{i\lambda
_{+}}\left| \Uparrow_{n} (t=0)\right\rangle =\text{e}^{i\gamma _{+}(\theta )}%
\text{e}^{i\lambda _{+}}\left| \Uparrow_{n} \right\rangle \\
\label{lamb1} \left| \Downarrow_{n} ;t=\tau \right\rangle
=\text{e}^{i\gamma _{-}(\theta )}\text{e}^{i\lambda
_{-}}\left| \Downarrow_{n} (t=0)\right\rangle =\text{e}^{i\gamma _{-}(\theta )}%
\text{e}^{i\lambda _{-}}\left| \Downarrow_{n} \right\rangle.
\end{eqnarray}
Here $\gamma_{+}$ is the Berry phase picked up in the magnetic
field along $\vec{n}(\theta)$, $\gamma_{-}$ is that along
$\vec{n}(\pi-\theta)$. To evaluate those Berry phases we consider
the poles at the degeneracy point $\mathbf{B}=0$.
\begin{equation}
\nabla _{\mathbf{B}(t^{\prime })}H(\mathbf{B}(t^{\prime })) =-\frac{1}{2}%
g\mu \sigma.
\end{equation}
\begin{equation}
 \gamma _{\pm }(\theta ) =-\frac{1}{2}\int\!\!\int_{S(C)}\frac{\mathbf{B}}{B^{2}}\cdot d%
\mathbf{B=}-\frac{1}{2}\int_{C}d\Omega =-\frac{1}{2}\Omega (\theta )
\end{equation}
\begin{eqnarray}
\gamma _{+} &=&-\pi (1-\cos \theta );\text{ \ for }\Omega (0\longrightarrow
\theta ) \\
\gamma _{-} &=&-\pi (1+\cos \theta );\text{ \ for }\Omega (0\longrightarrow
\pi -\theta )
\end{eqnarray}
then one has
\begin{equation}
\gamma _{+}(\theta )=-2\pi -\gamma _{-}(\theta )
\end{equation}
The global factor $\exp (iE_{0}t/\hslash )$ can be neglected for
normalization, and $\lambda_\pm$'s in Eq.(\ref{lamb1}) are set to be
\begin{equation}
\lambda _{+}=-2\pi \frac{\omega _{1}}{\omega _{0}}%
; \;\; \text {and }\;\;\; \lambda _{-}=2\pi \frac{\omega
_{1}}{\omega _{0}}.
\end{equation}
By making use of the above results  we can obtain the spin states
after the electron going through the two successive rotating
magnetic fields or the spin-echo device
\begin{equation}
\left| \Uparrow_{n} ;t=\tau \right\rangle _{1}=\text{e}^{i\gamma _{+}(\theta )}%
\text{e}^{i\lambda _{+}}\left| \Uparrow_{n} \right\rangle _{1}
\end{equation}
\begin{eqnarray}
\nonumber \left| \Downarrow_{n} ;t=\tau \right\rangle _{2}
&=&\text{e}^{i\gamma _{-}(\pi -\theta )}\text{e}^{i\lambda
_{-}}\left| \Downarrow_{n} ;t=0\right\rangle _{2} \\ \nonumber
&=&\text{e}^{i\gamma _{-}(\pi -\theta )}\text{e}^{i\lambda
_{-}}\left| \Uparrow_{n} ;t=\tau \right\rangle _{1} \\ \nonumber
&=&\text{e}^{i\gamma _{+}(\theta )}\text{e}^{i\lambda
_{-}}\text{e}^{i\gamma _{+}(\theta )}\text{e}^{i\lambda
_{+}}\left| \Uparrow_{n} ;t=0\right\rangle _{1}
\\
&=&\text{e}^{2i\gamma _{+}(\theta )}\left| \Uparrow_{n}
\right\rangle _{1}.
\end{eqnarray}
Here the suffix 1 or 2 refers the first or second magnetic
field. Similarly, one has
\begin{equation}
\left| \Downarrow_{n} ;t=\tau \right\rangle_{1} =\text{e}^{i\gamma _{-}(\theta )}%
\text{e}^{i\lambda _{-}}\left| \Downarrow_{n} ;t=0\right\rangle
_{1}
\end{equation}
\begin{eqnarray}
\nonumber \left| \Uparrow_{n} ;t=\tau \right\rangle _{2}
&=&\text{e}^{i\gamma _{+}(\pi -\theta )}\text{e}^{i\lambda
_{+}}\left| \Uparrow_{n} ;t=0\right\rangle _{2} \\ \nonumber
&=&\text{e}^{i\gamma _{+}(\pi -\theta )}\text{e}^{i\lambda
_{+}}\left| \Downarrow_{n} ;t=\tau \right\rangle _{1} \\ \nonumber
&=&\text{e}^{i\gamma _{-}(\theta )}\text{e}^{i\lambda
_{+}}\text{e}^{i\gamma _{-}(\theta )}\text{e}^{i\lambda
_{-}}\left| \Downarrow_{n} ;t=0\right\rangle
_{1} \\
&=&\text{e}^{2i\gamma _{-}(\theta )}\left| \Downarrow_{n}
\right\rangle _{1}
\end{eqnarray}
Consequently, an electron passing through the aforementioned two successive rotating
magnetic fields evolves to
\begin{eqnarray}
\left| \Uparrow_{n} \right\rangle \stackrel{2\tau }{\longrightarrow }\text{e}%
^{2i\gamma _{+}(\theta )}\left| \Uparrow_{n} \right\rangle
  \nonumber\\
\left| \Downarrow_{n} \right\rangle \stackrel{2\tau }{\longrightarrow }\text{e}%
^{2i\gamma _{-}(\theta )}\left| \Downarrow_{n} \right\rangle
\end{eqnarray}
Which can be written in a matrix form
\begin{equation}\label{eq:matrixform}
\pmatrix{a_{1\Uparrow_{n} }^{+}(2\tau ) \cr
         a_{1\Downarrow_{n} }^{+}(2\tau ) }
 =\pmatrix{
 \text{e}^{2i\gamma _{+}(\theta )} & 0 \cr
       0 & \text{e}^{2i\gamma _{-}(\theta )}
       }
 \pmatrix{
       a_{1\Uparrow_{n} }^{+}(0) \cr
       a_{1\Downarrow_{n} }^{+}(0)
       }
\end{equation}
where $a_{1\Uparrow_{n} }^{+}(2\tau)$ and
$a_{1\Downarrow_{n}}^{+}(2\tau )$ create electrons with spin in
direction of $\vec{n}(\theta,2\tau)$ and
$\vec{n}(\pi-\theta,2\tau)$, respectively, in channel 1 after a
period  $2\tau$. This gives the spin states $\left|\Uparrow_{n}
(t)\right\rangle$ and $\left|\Downarrow_{n} (t)\right\rangle$ at
the beam splitter after interacting with two successive opposite
rotating magnetic fields in channel 1. In terms of the inverse
relation of Eq.(\ref{zeign1}) and (\ref{zeign2}), one can
transform Eq.(\ref{eq:matrixform}) into the $\sigma_{z}$
representation,
\begin{equation}
\left(
\begin{tabular}{l}
$a_{1\uparrow }^{+}$ \\
$a_{1\downarrow }^{+}$%
\end{tabular}
\right)\!=\!\left(
\begin{tabular}{ll}
$\text{e}^{-2i\gamma _{+}(\theta )}\cos \frac{\theta }{2}$ & $-\text{e}%
^{-2i\gamma _{-}(\theta )}\sin \frac{\theta }{2}$ \\
$\text{e}^{-2i\gamma _{+}(\theta )}\sin \frac{\theta }{2}$ & $\text{e}%
^{-2i\gamma _{-}(\theta )}\cos \frac{\theta }{2}$%
\end{tabular}
\right)\! \left(
\begin{tabular}{l}
$a_{1\Uparrow_{n} }^{+}(2\tau )$ \\
$a_{1\Downarrow_{n} }^{+}(2\tau )$%
\end{tabular}
\right),
\end{equation}
where $a_{1\uparrow }^{+}$ and $a_{1\downarrow }^{+}$ create
electrons in the direction of $z$ and $-z$, respectively, in
channel 1 after a period $2\tau$. The spin states in channel 2
undergo a free motion before passing through the beam splitter. It
can be written as
\begin{equation}
\left(
\begin{tabular}{l}
$a_{2\uparrow }^{+}$ \\
$a_{2\downarrow }^{+}$%
\end{tabular}
\right) =\left(
\begin{tabular}{ll}
$\cos \frac{\theta }{2}$ & $-\sin \frac{\theta }{2}$ \\
$\sin \frac{\theta }{2}$ & $\cos \frac{\theta }{2}$%
\end{tabular}
\right) \left(
\begin{tabular}{l}
$a_{2\Uparrow }^{+}(2\tau)$ \\
$a_{2\Downarrow}^{+}(2\tau)$%
\end{tabular}
\right).
\end{equation}
At the beam splitter the electrons are scattered into outgoing channels 3 and 4.
The output can be expressed in terms of scattering matrix of
the beam splitter
\begin{equation}
\left(
\begin{tabular}{l}
$a_{3\Uparrow }^{+}$ \\
$a_{3\Downarrow }^{+}$ \\
$a_{4\Uparrow }^{+}$ \\
$a_{4\Downarrow }^{+}$%
\end{tabular}
\right) =\left(
\begin{tabular}{llll}
$r$ & $0$ & $t$ & $0$ \\
$0$ & $r$ & $0$ & $t$ \\
$t$ & $0$ & $r$ & $0$ \\
$0$ & $t$ & $0$ & $r$%
\end{tabular}
\right) \left(
\begin{tabular}{l}
$a_{1\Uparrow }^{+}(2\tau )$ \\
$a_{1\Downarrow }^{+}(2\tau )$ \\
$a_{2\Uparrow }^{+}(2\tau )$ \\
$a_{2\Downarrow }^{+}(2\tau )$%
\end{tabular}
\right),
\end{equation}
where $a_{3\Uparrow }^{+}$ and $a_{3\Downarrow }^{+}$ create
electrons with spin in direction of $\vec{n}(\theta,2\tau)$ and
$\vec{n}(\pi-\theta,2\tau)$, respectively, in channel 3. It is the
same for $a_{4\Uparrow }^{+}$ and $a_{4\Downarrow }^{+}$ in
channel 4. The combination of these two processes gives rise to
\begin{widetext}
\begin{equation}
\left(
\begin{tabular}{l}
$a_{3\uparrow }^{+}$ \\
$a_{3\downarrow }^{+}$ \\
$a_{4\uparrow }^{+}$ \\
$a_{4\downarrow }^{+}$%
\end{tabular}
\right) =\left(
\begin{tabular}{cccc}
$r(\cos2\gamma _{-}-i\cos\theta\sin2\gamma_{-})$ & $-ir\sin \theta \sin
(2\gamma _{-})$ & $t$ & $0$ \\
$-ir\sin \theta \sin (2\gamma _{-})$ & $r(\cos2\gamma _{-}+i\cos\theta\sin2\gamma_{-})$ & $0$ & $t$ \\
$t(\cos2\gamma _{-}-i\cos\theta\sin2\gamma_{-})$ & $-it\sin \theta \sin
(2\gamma _{-})$ & $r$ & $0$ \\
$-it\sin \theta \sin (2\gamma _{-})$ & $t(\cos2\gamma _{-}+i\cos\theta\sin2\gamma_{-})$ & $0$ & $r$%
\end{tabular}
\right) \left(
\begin{tabular}{l}
$a_{1\uparrow }^{+}$ \\
$a_{1\downarrow }^{+}$ \\
$a_{2\uparrow }^{+}$ \\
$a_{2\downarrow }^{+}$%
\end{tabular}
\right),  \label{sm}
\end{equation}
\end{widetext}
where $a_{3\uparrow }^{+}$ and $a_{3\downarrow }^{+}$ create
electrons in the direction of $z$ and $-z$, respectively, in
outgoing channel 3 after a period $2\tau$. The notation
implications for $a_{4\uparrow }^{+}$ and $a_{4\downarrow }^{+}$
are similar. In scattering matrix language it is
\begin{equation}
\left(
\begin{tabular}{l}
$a_{3\uparrow }$ \\
$a_{3\downarrow }$ \\
$a_{4\uparrow }$ \\
$a_{4\downarrow }$%
\end{tabular}
\right) =\left(
\begin{tabular}{llll}
$s_{3\uparrow ,1\uparrow }$ & $s_{3\uparrow ,1\downarrow }$ & $s_{3\uparrow
,2\uparrow }$ & $s_{3\uparrow ,2\downarrow }$ \\
$s_{3\downarrow ,1\uparrow }$ & $s_{3\downarrow ,1\downarrow }$ & $%
s_{3\downarrow ,2\uparrow }$ & $s_{3\downarrow ,2\downarrow }$ \\
$s_{4\uparrow ,1\uparrow }$ & $s_{4\uparrow ,1\downarrow }$ & $s_{4\uparrow
,2\uparrow }$ & $s_{4\uparrow ,2\downarrow }$ \\
$s_{4\downarrow ,1\uparrow }$ & $s_{4\downarrow ,1\downarrow }$ & $%
s_{4\downarrow ,2\uparrow }$ & $s_{4\downarrow ,2\downarrow }$%
\end{tabular}
\right) \left(
\begin{tabular}{l}
$a_{1\uparrow }$ \\
$a_{1\downarrow }$ \\
$a_{2\uparrow }$ \\
$a_{2\downarrow }$%
\end{tabular}
\right), \label{smat}
\end{equation}
where the s-matrix satisfies
\begin{equation}
S^{\dagger}S=SS^{\dagger}=I \label{su}
\end{equation}
and then the Eq.(\ref{smat}) can be written in a more compact form
\begin{equation}
a_{k\sigma }=\sum_{l \sigma ^{\prime }}s_{k\sigma ,l \sigma
^{\prime }}a_{l \sigma ^{\prime }};\text{ \ \ }k\in 3,4;\text{ \ }%
l \in 1,2;\text{ \ }\sigma ,\sigma ^{\prime }\in \uparrow
,\downarrow \label{asr1}.
\end{equation}
The creation and annihilation operators obey the conventional
communication relations
\begin{equation}
[a_{l\sigma
},a_{l^{\prime}\sigma ^{\prime }}^{+}]=\delta_{ll^{\prime}}\delta _{\sigma \sigma ^{\prime }}\text{ \ \ }%
l,l^{\prime} \in 1,2;\text{ \ }\sigma ,\sigma ^{\prime }\in
\uparrow ,\downarrow \label{asr2}
\end{equation}
and the outgoing operators also obey the communication relations
\begin{equation}
[a_{k\sigma
},a_{k^{\prime}\sigma ^{\prime }}^{+}]=\delta_{kk^{\prime}}\delta _{\sigma \sigma ^{\prime }}\text{ \ \ }%
k,k^{\prime} \in 3,4;\text{ \ }\sigma ,\sigma ^{\prime }\in
\uparrow ,\downarrow \label{asr3}.
\end{equation}
The incident operators $a_{l\sigma }$ and outgoing operators $a_{k\sigma ^{\prime }}^{+}$
satisfy the following relations
\begin{equation}
[a_{l\sigma },a_{k\sigma ^{\prime }}^{+}]=s_{k\sigma ^{\prime
},l\sigma }^{\ast}\text{ \ \ \ }l\in 1,2;\text{ \ \ }k\in 3,4
\label{asr4}
\end{equation}
With relations Eqs.(\ref{su}-\ref{asr4}), we are in the position
to calculate physical observable. In this paper we are interested
in the correlations of two electrons forming entangled states or
polarized states. There are four kind of states emitted from the
entangler. One is singlet with total spin zero and the others are
three triplets with total spin one. The scattering region involves
interaction with magnetic fields and the beam splitter. The
incident singlet state is denoted by $\left| \Psi
_{S}\right\rangle=1/\sqrt{2}(a_{1\uparrow }^{+}a_{2\downarrow
}^{+}-a_{1\downarrow }^{+}a_{2\uparrow }^{+})\left| 0\right\rangle
$ and the entangled triplet state is denoted by $\left| \Psi
_{T_{e}}\right\rangle=1/\sqrt{2}(a_{1\uparrow }^{+}a_{2\downarrow
}^{+}+a_{1\downarrow }^{+}a_{2\uparrow }^{+})\left| 0\right\rangle
$. The other two unentangled triplet states are $\left| \Psi
_{T_{u}}\right\rangle=a_{1\uparrow }^{+}a_{2\uparrow }^{+}\left|
0\right\rangle  $ and $\left| \Psi
_{T_{d}}\right\rangle=a_{1\downarrow }^{+}a_{2\downarrow
}^{+}\left| 0\right\rangle  $ for spins in parallel upward and
downward in $z$ axis respectively. The suffix $1$ refers for
channel $1$ and the suffix $2$ for channel $2$ as shown in Fig. 1.
The two electrons emitted from the entangler are spatially
separated in each channel due to the Coulomb blockade occurring in
the quantum dots\cite{recher,buttiker1}. The electron in channel
$1$ is scattered by the rotating magnetic fields picking up a
geometric phase and  then moves to the beam splitter. The two
electrons interact at the beam splitter and move to the output
channel 3 or 4. The outgoing amplitudes of electron waves in probe
$3$ and $4$ are determined by Eq.(\ref{sm}), i.e., $\left
|\text{out}\right\rangle=S^{\dagger}\left
|\text{in}\right\rangle$. In terms of S-matrix the incoming states
are related to the outgoing states by $\left
|\text{in}\right\rangle=S\left |\text{out}\right\rangle$ due to
the unitary condition of s-matrix given in Eq.(\ref{su}).

The incoming states can be conveniently expressed by outgoing states in operator form
\begin{widetext}
\begin{eqnarray} \nonumber
\left\{
\begin{array}{c}
\left| \Psi _{S}\right\rangle  \\
\left| \Psi _{T_{e}}\right\rangle
\end{array}
\right.&=&\frac{1}{\sqrt{2}}(a_{1\uparrow }^{+}a_{2\downarrow
}^{+}\mp
a_{1\downarrow }^{+}a_{2\uparrow }^{+})\left| 0\right\rangle  \\
&=&\frac{1}{\sqrt{2}}\left [\left\{
\begin{array}{c}
P \\
\tilde{P}
\end{array}
\right. (a_{3\uparrow }^{+}a_{3\downarrow }^{+}+a_{4\uparrow
}^{+}a_{4\downarrow }^{+})+\left\{
\begin{array}{c}
Q \\
\tilde{Q}
\end{array}
\right. a_{3\uparrow }^{+}a_{4\downarrow }^{+}+\left\{
\begin{array}{c}
M \\
\tilde{M}
\end{array}
\right. a_{3\downarrow }^{+}a_{4\uparrow }^{+}+\left\{
\begin{array}{c}
D \\
\tilde{D}
\end{array}
\right. (a_{3\uparrow }^{+}a_{4\uparrow }^{+}+a_{3\downarrow
}^{+}a_{4\downarrow }^{+}) \right ] \left |0\right \rangle
\label{inout1}
\end{eqnarray}

\begin{equation}
\left| \Psi _{u}\right\rangle =a_{1\uparrow }^{+}a_{2\uparrow
}^{+}\left| 0\right\rangle =\left [P_{u}a_{3\uparrow
}^{+}a_{4\uparrow }^{+}+Q_{u}(a_{3\downarrow }^{+}a_{3\uparrow
}^{+}+a_{4\downarrow }^{+}a_{4\uparrow }^{+})+S_{u}(r^{\ast
2}a_{3\downarrow }^{+}a_{4\uparrow }^{+}-t^{\ast 2}a_{3\uparrow
}^{+}a_{4\downarrow }^{+}) \right ] \left |0\right \rangle
\label{inout2}
\end{equation}
\begin{equation}
\left| \Psi _{d}\right\rangle =a_{1\downarrow }^{+}a_{2\downarrow
}^{+}\left| 0\right\rangle=\left [P_{d}a_{3\downarrow
}^{+}a_{4\downarrow }^{+}+Q_{d}(a_{3\uparrow }^{+}a_{3\downarrow
}^{+}+a_{4\uparrow }^{+}a_{4\downarrow }^{+})+S_{d}(r^{\ast
2}a_{3\uparrow }^{+}a_{4\downarrow }^{+}-t^{\ast 2}a_{3\downarrow
}^{+}a_{4\uparrow }^{+}) \right ] \left |0 \right \rangle
\label{inout3}
\end{equation}
\end{widetext}
where
\begin{eqnarray} \nonumber
P&=&2r^{\ast }t^{\ast }\cos 2\gamma _{-}; \ \ \tilde{P}=2ir^{\ast
}t^{\ast }\cos \theta \sin 2\gamma _{-}\\ \nonumber Q&=&(r^{\ast
2}+t^{\ast 2})\cos 2\gamma _{-}+i(r^{\ast 2}-t^{\ast 2})\cos
\theta \sin 2\gamma _{-} \\ \nonumber \tilde{Q}&=&(r^{\ast
2}-t^{\ast 2})\cos 2\gamma _{-}+i(r^{\ast 2}+t^{\ast 2})\cos
\theta \sin 2\gamma _{-} \\ \nonumber M&=&-[(r^{\ast 2}+t^{\ast
2})\cos 2\gamma _{-}+i(t^{\ast 2}-r^{\ast 2})\cos \theta \sin
2\gamma ]\\ \nonumber \tilde{M}&=&(r^{\ast 2}-t^{\ast 2})\cos
2\gamma _{-}-i(r^{\ast 2}+t^{\ast 2})\cos \theta \sin 2\gamma _{-}
\\ \nonumber
D&=&i\sin \theta \sin 2\gamma _{-}(r^{\ast 2}-t^{\ast 2}) \\
\nonumber \tilde{D}&=&i\sin \theta \sin 2\gamma _{-}(r^{\ast
2}+t^{\ast 2}) \\ \nonumber P_{u}&=&(\cos 2\gamma _{-}+i\sin
2\gamma _{-}\cos \theta )(r^{\ast 2}-t^{\ast 2}) \\ \nonumber
P_{d}&=&(\cos 2\gamma _{-}-i\sin 2\gamma _{-}\cos \theta )(r^{\ast
2}-t^{\ast 2}) \\ \nonumber Q_{u}&=&ir^{\ast }t^{\ast }\sin \theta
\sin (2\gamma _{-}) \\ \nonumber Q_{d}&=&ir^{\ast }t^{\ast }\sin
\theta \sin (2\gamma _{-}) \\ \nonumber S_{u}&=&i\sin \theta \sin
(2\gamma _{-}) \\ \nonumber S_{d}&=&i\sin \theta \sin (2\gamma
_{-}) \nonumber.
\end{eqnarray}
The probabilities of the electron with spin $\sigma$ appearing at
output probes $3$ and $4$ can be evaluated by
\begin{equation}
\left\langle \Psi _{j}\right| n_{l\sigma }\left| \Psi
_{j}\right\rangle \ \ \text{or} \ \ \ \left\langle \Psi
_{j}\right| n_{l\sigma }n_{l^{\prime }\sigma ^{\prime }}\left|
\Psi _{j}\right\rangle,
\end{equation}
where $j$ is label for the singlet and triplets, and $l,$
$l^{\prime }\in 3,$ $4$, $\sigma ,$
$\sigma ^{\prime }\in \uparrow ,$ $\downarrow $. For instance, $\ j=T_{e}$ and $%
\ l=3,$ $l^{\prime }=4$. $\sigma =\uparrow $ and $\sigma ^{\prime
}=\downarrow $. One has to evaluate the expectation that one
electron appears at the output probe 3 with spin up and the other
electron appears at the output probe 4 with spin down
simultaneously for the incident state $\left |\Psi
_{T_{e}}\right\rangle$.
\begin{eqnarray} \nonumber
&&\left\langle \Psi _{T_{e}}\right|n_{3\uparrow }n_{4\downarrow
}(1-n_{3\downarrow})(1-n_{4\uparrow})\left| \Psi
_{T_{e}}\right\rangle\\ \nonumber &=&\left\langle \Psi
_{T_{e}}\right| a_{3\uparrow }^{+}a_{3\uparrow }a_{4\downarrow
}^{+}a_{4\downarrow }(1-a_{3\downarrow }^{+}a_{3\downarrow
})(1-a_{4\uparrow }^{+}a_{4\uparrow })\left| \Psi
_{T_{e}}\right\rangle
\\ \nonumber &=&\left\langle \Psi _{T_{e}}\right| a_{3\uparrow
}^{+}a_{4\downarrow }^{+}a_{4\downarrow }a_{3\uparrow }\left| \Psi
_{T_{e}}\right\rangle
\\ \nonumber &=&\sum_{m}\left\langle \Psi _{T_{e}}\right| a_{3\uparrow
}^{+}a_{4\downarrow }^{+}\left| m\right\rangle \left\langle
m\right| a_{4\downarrow }a_{3\uparrow }\left| \Psi
_{T_{e}}\right\rangle  \\ \nonumber &=&\sum_{m}|\left\langle
m\right| a_{4\downarrow }a_{3\uparrow }\left| \Psi
_{T_{e}}\right\rangle |^{2}
\\ \nonumber
&=&\frac{1}{2}\sum_{m}|\left\langle m\right| a_{4\downarrow
}a_{3\uparrow }(a_{1\uparrow }^{+}a_{2\downarrow
}^{+}-a_{1\downarrow }^{+}a_{2\uparrow }^{+})\left| 0\right\rangle |^{2},
\label{exap}
\end{eqnarray}
where $m$ runs over all the states that constitute a complete set
of Hilbert space. We have dropped terms involving more than three
annihilation operators, such as $a_{3\uparrow
}a_{3\downarrow}a_{4\downarrow}$ due to the Pauli exclusive
principle. Using Eqs.(\ref{inout1})-(\ref{inout3}) and commutation
relations of output operators $a^{+}_{3\sigma}$ and
$a^{+}_{4\sigma}$, one can obtain the output mean electron
numbers, variances and correlations for the input states, {\it
i.e.}, the singlet and triplet states. The results are summarized
in the TABLE I and TABLE II. The TABLE III is the input mean
numbers, variance and correlations for the four incident states.

\section{Results and discussions}

TABLE I gives
 the output correlations and average numbers of electrons and
 TABLE II gives output variances of electrons for input
 singlet and triplet states. As a comparison we list input correlations,
 average occupations and variances of electrons for input singlet
 and triplet states in TABLE III.
Our discussions contain the following parts

 \subsection{The average occupation in one output channel}
\begin{figure}[tbph]
\vbox to 2.5 in{\rule{0pt}{1in}} \includegraphics{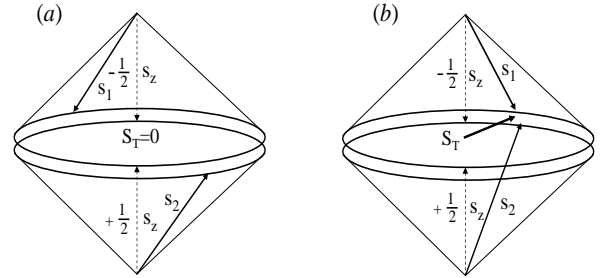}
\caption{The composition of two electron spins. (a) the singlet
state composed of two opposite spins $s_{1}$ and $s_{2}$ and their
$z$ components. The total spin $S_{T}$ and the total $z$ component
are zero. (b) The triplet with its zero $z$ component. The two
spins $s_{1}$ and $s_{2}$ are not opposite and form a total spin
$S_{T}=1$. } \label{fig2}
\end{figure}

\begin{figure}[]
\vbox to 2in{\rule{0pt}{1.8in}} \includegraphics{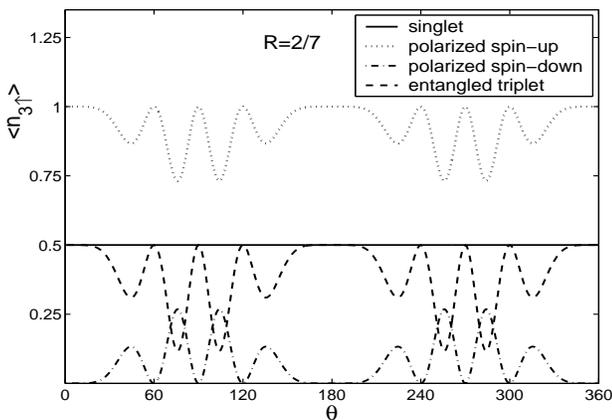}
\caption{The average occupation probabilities of the spin-up in
the output channel 3 for different incident states. The constant
value $0.5$ is for the incident singlet state and shown by the
solid line. The dashed line is for the incident entangled triplet
state. The dotted line is for the incident spin-up polarized state
and the dash-dotted line is for the incident spin-down polarized
state. The beam-splitter is asymmetric with the reflection
probability $R=2/7.$
 }\label{fig3}
\end{figure}

\begin{figure}[]
\vbox to 2in{\rule{0pt}{1.8in}} \includegraphics{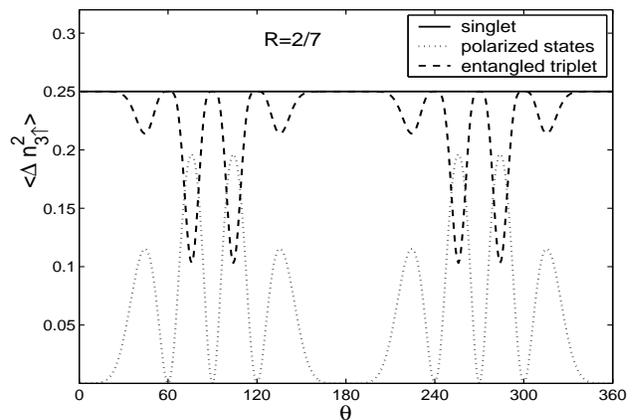}
\caption{The variances of the spin-up in the output channel 3 for
different incident states. The constant value $0.25$ is for the
incident singlet state and shown by the solid line. The dashed
line is for the incident entangled triplet state. The dotted line
is for both the incident  spin-up and spin-down polarized states.
The beam-splitter is the same as that in Fig.~\ref{fig3}.
}\label{fig4}
\end{figure}

 Firstly one can see that in the output
 channels the average numbers of electrons are constant values 0.5 for
 the entangled singlet states whether the entangled singlet state is scattered or not
 (see TABLE I and TABLE III rows 2 to 5 and column 2).
 They are irrelevant to magnetic field and the beam splitter. However, for the incident entangled
 triplet state the output average numbers of
 electrons  are dependent on Berry phase
 (see TABLE I rows 2 to 5 and column 3).
 In comparison to the initial average  numbers of electrons
 (see TABLE III rows 2 to 5 and column 3), the scattered entangled triplet state
 acquires additional phase.  These results indicate that the singlet state does not acquire
 a phase in the rotating magnetic fields. It is just
 like a composite particle with zero spin, as shown by Fig.~\ref{fig2}a.
 In Fig.~\ref{fig2}a the two spins are always aligned opposite direction.
 They form a composite particle with the total
 spin zero $S_{T}=0$. In this case the $z$ components of the two electrons are
 opposite in direction and the composite particle is just
 the singlet.  With this notation in mind
 the entangled triplet state should acquire Berry phase due to its
 total spin $S_{T}=1$,  although the total projection of spin along $z$ direction is zero,
 (see TABLE I rows 2 to 5 and column 3 and Fig.~\ref{fig2}b).
 The average output numbers of electrons are related to Berry phase
 for the input polarized triplet states. But they are different
 from that of the entangled triplet state.
 For example, in the input polarized state $\left
|\Psi_{T_{u}}\right\rangle=a^{+}_{1\uparrow}a^{+}_{2\uparrow}\left
|0\right\rangle=\left |\uparrow \right\rangle_{1}\otimes\left
|\uparrow\right\rangle_{2}$ the electron in channel 1 with spin-up
subjects the rotating magnetic fields and is scattered at the beam
splitter. The reduced part of $\left\langle\Psi_{T_{u}}|
n_{3\uparrow}|\Psi_{T_{u}}\right\rangle$ is $R\Theta$. This is
just the reflected spin flipped part of spin-up in channel 1 into
spin-down. Viewing
$\left\langle\Psi_{T_{u}}|n_{3\downarrow}|\Psi_{T_{u}}\right\rangle=R\Theta$
one finds that the contribution of the spin flipped part is
precisely  $R\Theta$. Similar analysis shows that  the spin
flipped part is $T\Theta$ for channel 1 electrons, which transmit
to the output channel 4. It can also be seen clearly by setting
the total reflection beam splitter, i.e., $R=1$. In this condition
the electron in channel 1 directly goes to the output channel 3
and the electron in channel 2 directly goes to the output channel
4. Because the rotating magnetic fields are located in the channel
1, the transmission of an electron from the channel 2 to the
channel 4 is unit for the spin conserved processes (i.e., the
spin-up to spin-up and spin-down to spin-down) and  zero for spin
flipped processes (i.e., spin-up to spin-down and spin-down to
spin-up). It is verified in TABLE I rows 4 and 5, columns 4 and 5
by setting $T=0$. Fig.~\ref{fig3} and Fig.~\ref{fig4} show the
variations of the occupation probability and its mean square
fluctuation of the spin-up electrons in the output channel 3 with
the angle $\theta$, which is the polar angle of the total magnetic
field along $z$-axis, as shown in Fig.~\ref{fig1}. The beam
splitter is asymmetric by taking reflection coefficient $R=2/7$.
From Fig.~\ref{fig3} and Fig.~\ref{fig4} one can see that the
averages $\left\langle n_{3\uparrow}\right\rangle$ are different
functions of $\theta$ for each incident state, which can be
controlled by $B_{z}$ due to $\tan\theta=B_{\perp}/B_{z}$.
$B_{\perp}$ is a rotating magnetic field in $x$-$y$ plane whose
magnitude is fixed. $B_{z}$ is a constant magnetic field along $z$
axis which can be controlled by the external means.  From above
results and discussions it can motivate us to use a counter
detector to distinguish the four states by varying the magnetic
field $B_{z}$.

\subsection{The same output channel with different spins}
\begin{figure}[tbph]
\vbox to 2.2in{\rule{0pt}{1.8in}} \includegraphics{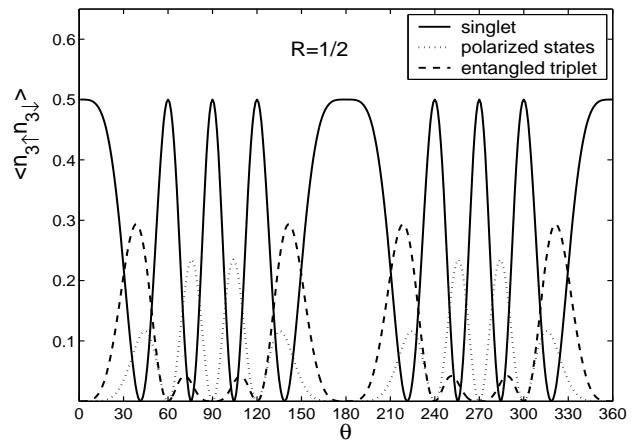}
\caption{The correlations of the spin-up and spin-down in the same
output channel 3 for different incident states. The solid line is
for the singlet state. The dashed line is for the incident
entangled triplet state. The dotted line is for both the incident
spin-up and spin-down polarized states. The beam-splitter is an
ideal device with the ratio of reflection to transmission
50:50.}\label{fig5}
\end{figure}

\begin{figure}[tbph]
\vbox to 2.2in{\rule{0pt}{1.8in}} \includegraphics{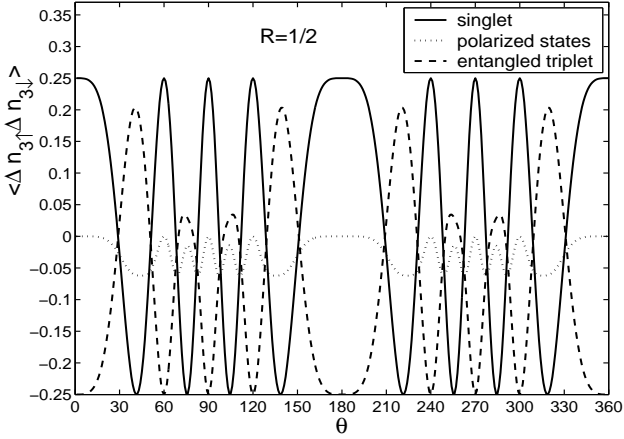}
\caption{The variances of the spin-up and spin-down in the same
output channel 3 for different incident states. The solid line is
for the entangled singlet state. The dashed line is for the
incident entangled triplet state. The dotted line is for both the
incident spin-up and spin-down polarized states. The beam-splitter
is the same as Fig.~\ref{fig5}.}\label{fig6}
\end{figure}

  From rows 6 to 7  in TABLE I and TABLE II, one can see that the
  correlations $\left\langle n_{3\uparrow}n_{3\downarrow}\right\rangle$ and
  $\left\langle n_{4\uparrow}n_{4\downarrow}\right\rangle$ are different for entangled
  singlet and triplet states, whereas the same for unentangled states.
  In the entangled singlet state the correlation of
  $\left\langle\Psi_{S}\left|n_{3\uparrow}n_{3\downarrow}\right|\Psi_{S}\right\rangle$
  reaches the maximum $2RT$ and minimum 0 , by viewing a cosine function in the
  expression
  $\left\langle\Psi_{S}\left|n_{3\uparrow}n_{3\downarrow}
  \right|\Psi_{S}\right\rangle=2RT\cos^{2}2\gamma_{-}$.
  The correlations oscillate with respect to the angle $\theta$, as shown by solid line in Fig.~\ref{fig5}.
  In Fig.~\ref{fig5} the dashed line represents the correlation for the input entangled triplet. One
  can see that it oscillates but the amplitude is less than that of the singlet. The dotted line is for two polarized
  states. Fig.~\ref{fig6} shows the variances
  $\left\langle\Delta n_{3\uparrow}\Delta n_{3\downarrow}\right\rangle$ for incident states.
  It shows that the variance of the input singlet state oscillates
  positively and negatively.

When the variance is positive the two electrons prefer to reach the same output channel,
bunching behavior.
When the variance is negative, they prefer to separate to different output channels,
antibunching behavior.
   Due to Berry phase picked up by spins passing
  through the rotating magnetic fields, the symmetry of the wave
  function is changed. The correlation
  $\left\langle n_{3\uparrow}n_{3\downarrow}\right\rangle$ goes to zero at
  $\theta=41.4^{\circ}, 75.5^{\circ}, 104.5^{\circ}$ and
  $138.6^{\circ}$,
  and reaches the maximum at $\theta=0^{\circ}, 60^{\circ}, 90^{\circ},
  120^{\circ}$ and $180^{\circ}$ for the singlet state and an ideal beam splitter, $R=T=1/2$.
  Correspondingly, the variances $\left\langle \Delta n_{3\uparrow}\Delta
  n_{3\downarrow}\right\rangle$ reach maximal positive and negative
  values at these angles for the singlet state.
  For the entangled triplet the situation
  is reversed. At $\theta=41.4^{\circ}, 75.5^{\circ}, 104.5^{\circ}$ and
  $138.6^{\circ}$ the correlations are maximum and
  at $\theta=0^{\circ}, 60^{\circ}, 90^{\circ},
  120^{\circ}$ and $180^{\circ}$ are minimum.
  The variances $\left\langle\Delta n_{3\uparrow}\Delta n_{3\downarrow}\right\rangle$ are always negative
  showing the antibunching property for the polarized states, as shown by dotted line in
  Fig.~\ref{fig6}.
  The same analysis is applied for the output channel 4.\\

\subsection{The output correlations of different  channels with the same spins}

\begin{figure}[]
\vbox to 2.2in{\rule{0pt}{1.8in}} \includegraphics{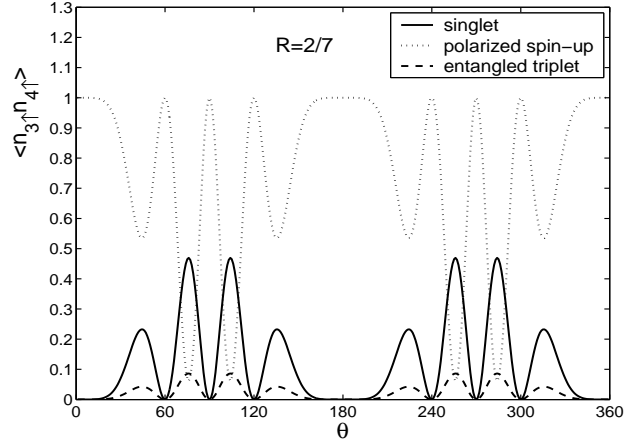}
\caption{The correlations of spin-up and spin-up in different
output channels 3 and 4 for the unsymmetrical beam splitter. The
reflection of the beam splitter is $R=2/7$. The solid line is for
singlet and the dashed line is for the entangled triplet. The
dotted line is for  spin-up polarized state. The correlation of
the spin-down polarized state is constant zero. }\label{fig7}
\end{figure}

\begin{figure}[]
\vbox to 2.2in{\rule{0pt}{1.8in}} \includegraphics{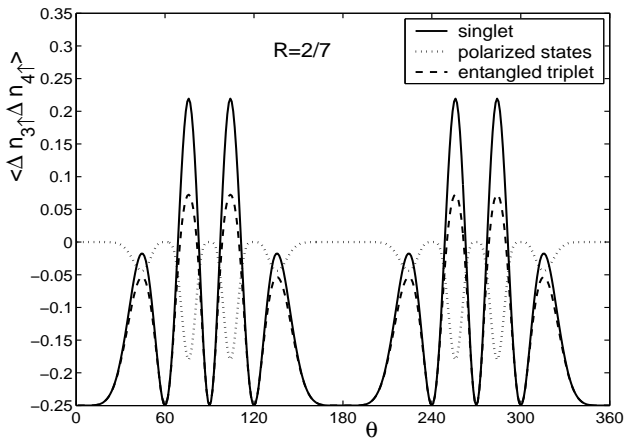}
\caption{The variances of the two spin-up electrons in different
output channels 3 and 4. The reflection of the beam splitter is
$R=2/7$. The solid line is for the singlet, the dashed line is for
the entangled triplet. The dotted line is for both spin-up and
spin-down polarized states. }\label{fig8}
\end{figure}

The correlations $\left\langle n_{3\uparrow}n_{4\uparrow}\right\rangle$
  and $\left\langle n_{3\downarrow}n_{4\downarrow}\right\rangle$ are
  different for the entangled  singlet and triplet states.
  They also differ from that of unentangled triplet states $\left
  |\Psi_{T_{u}}\right\rangle$ and $\left
  |\Psi_{T_{d}}\right\rangle$. From TABLE I and TABLE II rows 8 to 9
  one can find that the correlation is definite 0 for the input state $\left
  |\Psi_{T_{d}}\right\rangle$. This is due to the particle property of an electron.
  The particle occurs only at one position at a time, although there is a probability
  to flip a spin in channel 1 for the incident spin-down state $\left
  |\Psi_{T_{d}}\right\rangle=a_{1\downarrow}^{+}a_{2\downarrow}^{+}\left|0\right\rangle$
  splitting into the output channel 3 and 4.
  The particle cannot be partially reflected
  to channel 3 and partially transmitted to channel 4
  simultaneously. If the magnetic fields is removed,
  the correlations become 0 for the entangled states
  and 1 for the spin-up polarized state.
  Under this condition the variance
  $\left\langle\Delta n_{3\uparrow}\Delta n_{4\uparrow}\right\rangle$
  is 0 for untangled states and $-1/4$ for the entangled states.
  These results manifest  that the electrons with
   spin-up in channel 3 and 4 cannot appear simultaneously
   in the input state $\left|\Psi_{T_{d}}\right\rangle$.
   In addition the fluctuation
   $\left\langle\Delta n_{3\uparrow} \Delta n_{4\uparrow}\right\rangle$
   in this case vanishes due to the Pauli exclusive principle in the
   input state $\left|\Psi_{T_{u}}\right\rangle$.
   It is interesting to note that the correlation $n_{3\uparrow}$ and $n_{4\uparrow}$
   of the input entangled triplet state is zero when the beam splitter is ideal,
   {\it i.e.}, 50\% reflection and 50\% transmission.
   In this case the variance is definitely negative, as shown by the
   expression in TABLE II (rows 8 and 9, column 3).
   These negative values indicate that
   the two electrons with the same spin prefer to locate in different
   output channels simultaneously in entangled singlet and triplet states.
One can see that the correlations $\left\langle
n_{3\uparrow}n_{4\uparrow}\right\rangle$ are different in all
incident states for asymmetric beam splitter (shown by
Fig.~\ref{fig7} and Fig.~\ref{fig8}). In this case the two
correlations and variances of the entangled states behave
similarly (shown by solid line and dashed line in Fig.~\ref{fig7},
\ref{fig8}, respectively). This again indicates that the output
signals $\left\langle n_{3\uparrow}n_{4\uparrow}\right\rangle$ and
$\left\langle\Delta n_{3\uparrow}\Delta
n_{4\uparrow}\right\rangle$ are different with respect to Berry
phase.


\subsection{The output correlations of different channels with different
spins}

\begin{figure}[tbph]
\vbox to 2.1in{\rule{0pt}{1.8in}} \includegraphics{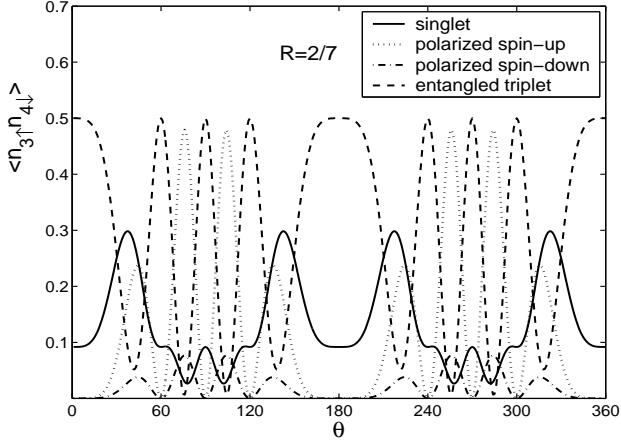}
\caption{The correlations of two electrons with opposite spins,
i.e., spin-up in  channel 3 and spin-down in channel 4. The beam
splitter is unsymmetrical with $R=2/7$. The solid line is for the
singlet and the dashed line is for the entangled triplet. The
dotted line is for spin-up polarized state and the dot-dashed lin
is for spin-down polarized state.}\label{fig9}
\end{figure}

\begin{figure}[tbph]
\vbox to 2.1in{\rule{0pt}{1.8in}} \includegraphics{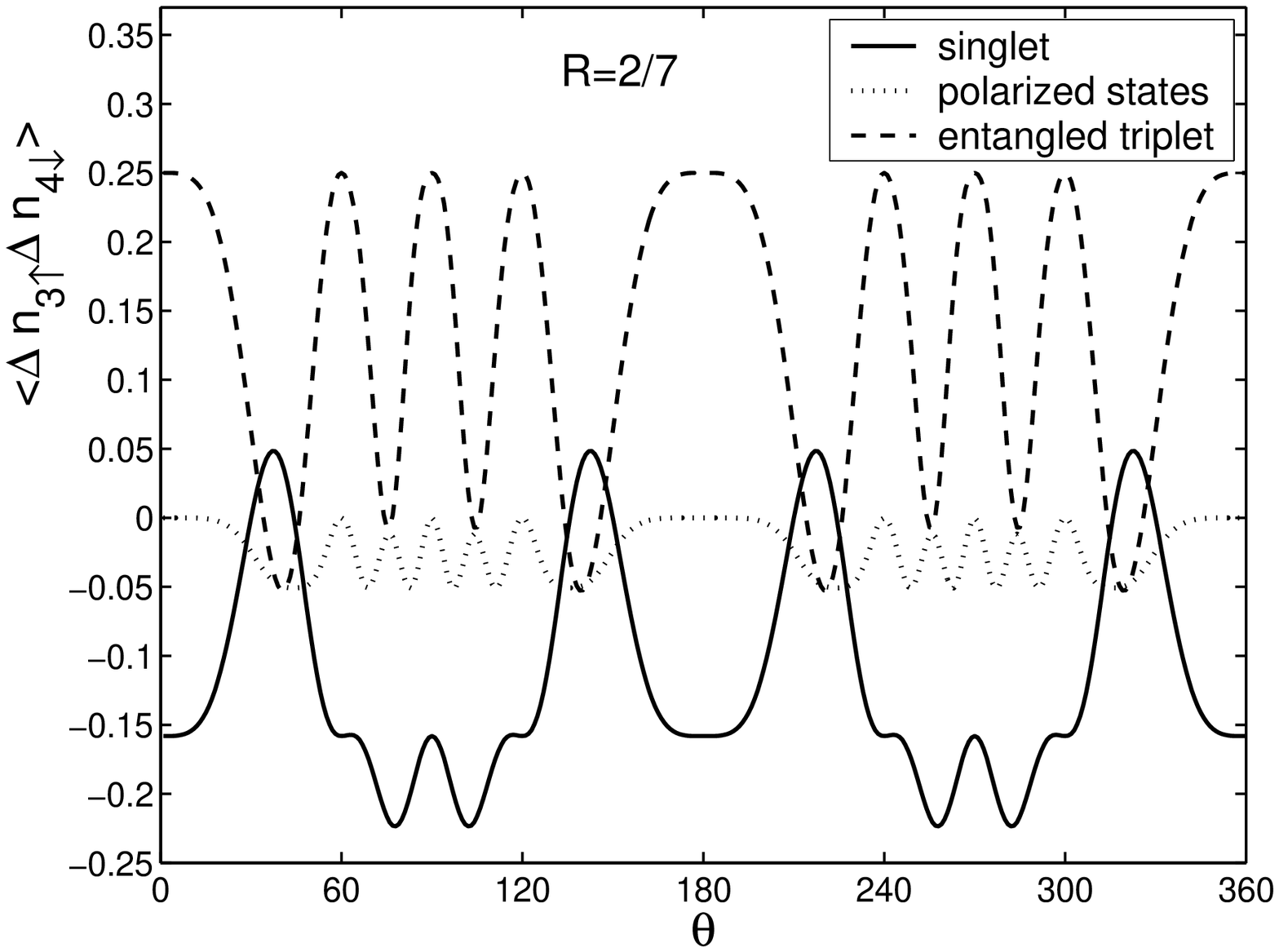}
\caption{The variances of the two electrons with spin-up in
channel 3 and spin-down in channel 4. The beam splitter is the
same as Fig.~\ref{fig9}. The solid line is for the singlet and the
dashed line is for the entangled triplet. The dotted line is for
both polarized states. }\label{fig10}
\end{figure}

\begin{figure}[tbph]
\vbox to 2.1in{\rule{0pt}{1.8in}} \includegraphics{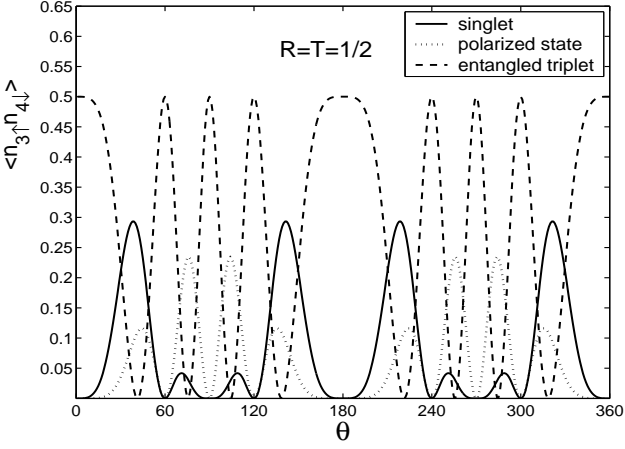}
\caption{The correlations of two electrons with opposite spins in
different channels 3 and 4. The beam splitter is ideal, i.e.
50:50. The solid line is for the singlet and the dashed line is
for the entangled triplet. The dotted line is for both polarized
states.}\label{fig11}
\end{figure}

\begin{figure}[tbph]
\vbox to 2.1in{\rule{0pt}{1.8in}} \includegraphics{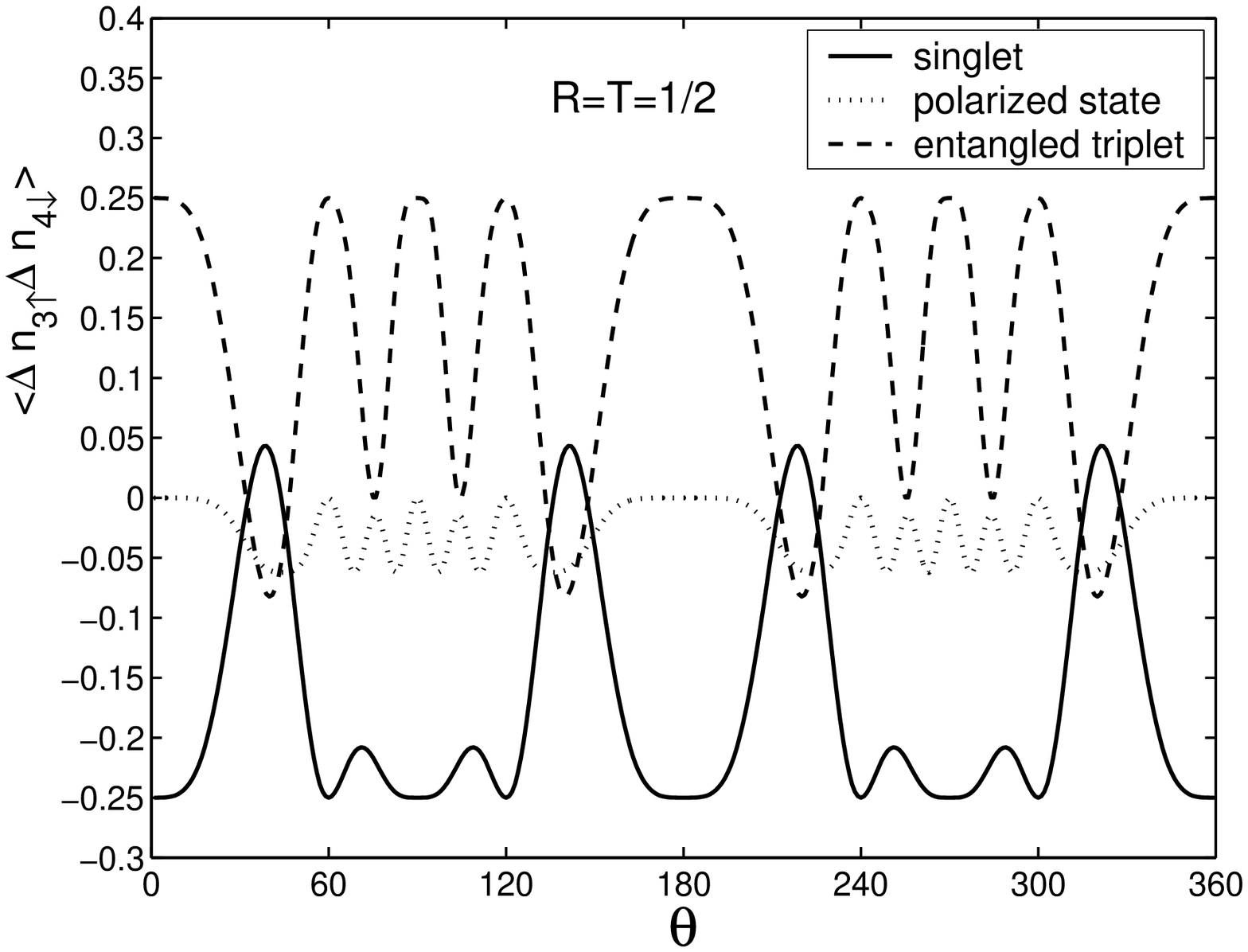}
\caption{The variances of the two electrons with spin-up in
channel 3 and spin-down in channel 4. The beam splitter is 50:50.
The solid line is for the singlet and the dashed line is for the
entangled triplet. The dotted line is for both polarized states.
}\label{fig12}
\end{figure}

 The rows 10 and 11 in TABLE I and TABEL II
give the correlations $\left\langle
n_{3\uparrow}n_{4\downarrow}\right\rangle$ and $\left\langle
n_{3\downarrow}n_{4\uparrow}\right\rangle$ for the incoming
singlet and triplet states. From those expressions one can see
that the four correlations are different in functions of Berry
phase, in the transmission, and in the reflection coefficients.
This is again possible to devise a coincident apparatus to
distinguish these four states by means of rotating magnetic fields
and the external gate voltages. Since the beam splitter is usually
constructed by a quantum point contact (QPC)\cite{liu}. The
transmission and reflection are adjusted by the potential barrier,
which is controlled by external gate voltages. If the beam
splitter is asymmetric, say $R=2/7$ as shown in Fig.~\ref{fig9}
and Fig.~\ref{fig10}, the four incident states manifest  different
behaviors with the polar angle $\theta$. In experiments these
coincident signals are easily distinguished. Although the
coincidences of polarized states are different, the variances of
these states are the same (see Fig.~\ref{fig10}). In
Fig.~\ref{fig10} the fluctuation correlation is negative for the
singlet state and positive for the entangled triplet state showing
again the bunching and antibunching property. For a 50:50 beam
splitter the singlet and the entangled triplet are irrelevant to
transmission and reflection coefficients of the beam splitter. The
maximum value of correlation is about 0.3 for the singlet state
and 0.5 for the entangled triplet, as shown by Fig.\ref{fig11}.
The variance $\left\langle \Delta n_{3\uparrow}\Delta
n_{4\downarrow}\right\rangle$ is almost negative for the singlet
state and positive for the entangled triplet, as shown by
Fig.\ref{fig12}. This shows an electron with spin-up in channel 3
cannot appear simultaneously with an electron with spin-down in
channel 4. But it is not true for the entangled triplet. The value
of correlation $\left\langle
n_{3\uparrow}n_{4\downarrow}\right\rangle$ can reach the maximum
$1/2$, which shows an electron with spin-up in channel 3 appears
simultaneously with an electron with spin-down in channel 4 with a
probability $1/2$. The other case $\left\langle
n_{3\downarrow}n_{4\uparrow}\right\rangle$ is also $1/2$. The sum
of these two case is 1. It indicates that these two cases occur
definitely and showing the antibunching property. It is not the
case for the two polarized states, although their correlations are
zero. The reason is that for an individual events $n_{3\uparrow}$
and $n_{4\downarrow}$ are not zero for the two entangled states
but $n_{4\downarrow}$ is zero for $\left
|\Psi_{T_{u}}\right\rangle$  and $n_{3\uparrow}$ is zero for
$\left |\Psi_{T_{d}}\right\rangle$.

One can recover the results in Refs.~\cite{loudon,rep} by removing
magnetic fields and sum up the spin indices. The singlet state
$\left| \Psi _{s}\right\rangle$ shows bunching property
$\left\langle n_{3}n_{4}\right\rangle=(T-R)^{2}$ and the triplet
state $\left| \Psi _{T_{e}}\right\rangle$ shows antibunching
property $\left\langle n_{3}n_{4}\right\rangle=1$.  The average
number is 1 for each output channel, i.e., $\left\langle
n_{3}\right\rangle=\left\langle n_{4}\right\rangle=1$. It is also
consistent with the results of Ref.~\cite{rep} that the
two-particle-occupation probability in one channel is P(2,0)=2RT
for Bosons and 0 for Fermions. In our case this corresponds to
$\left\langle n_{3\uparrow}n_{3\downarrow}\right\rangle=2RT$ for
the singlet state and $\left\langle
n_{3\uparrow}n_{3\downarrow}\right\rangle=0$ for the triplet state
$\left| \Psi _{T_{u}}\right\rangle$. However, the four states
manifest different behavior in magnetic fields.    For example, in
the case of an ideal beam splitter(T=R=1/2) the correlations are
$\left\langle
n_{3\uparrow}n_{4\downarrow}\right\rangle=0.5\cos^{2}\theta
\sin^{2}2\gamma_{-}$ for singlet state $\left|
\Psi_{S}\right\rangle$ and $\left\langle
n_{3\uparrow}n_{4\downarrow}\right\rangle=0.5\cos^{2}2\gamma_{-}$
for the triplet state $\left| \Psi_{T_{e}}\right\rangle$,
respectively. It is the same for the two polarized states $\left|
\Psi_{T_{u}}\right\rangle$ and $\left| \Psi_{T_{d}}\right\rangle$,
i.e., $\left\langle
n_{3\uparrow}n_{4\downarrow}\right\rangle=0.25\sin^{2}\theta
\sin^{2}2\gamma_{-}$. If the beam splitter is  not symmetric the
correlations $\left\langle
n_{3\uparrow}n_{4\downarrow}\right\rangle$ are different for these
four states, which can be seen from the expressions in TABLE I row
10 or 11.\\

\section{Conclusion}

We have investigated the correlations and fluctuations of
two-electron states ({\it i.e.}, singlet and triplet states )
affected by two oppositely rotating magnetic fields. We found that
the four states pick up Berry phases in different manner
manifesting different behaviors in the output correlations and
variances. The entangled singlet state shows bunching and
antibunching behaviors with respect to Berry phase. The entangled
triplet state also shows bunching and antibunching behaviors that
differ from that for the singlet state and polarized triplet
states. The significant result is that two entangled electrons
behave like a composed quasi-particle with total spin zero and
unit for the singlet state and the entangled triplet state,
respectively. This quasi-particle property is clearly illustrated
by Berry phase picked up by the entangled states. Berry phase
acquired by the unentangled states shows different behaviors from
that of the entangled states. Additionally, the setup discussed in
this paper is expected to investigate Berry phase by measuring the
correlation and fluctuation using the coincident techniques.


\begin{center}
\begin{table*}[htbp]
\caption{The output correlations and average number of electrons
for input singlet and triplets
$(\Theta=\sin^{2}\theta\sin^{2}2\gamma_{-})$} \label{tstla1}
\begin{tabular}{|l|c|c|c|c|}  \hline\hline
 & $\left| \Psi _{S}\right\rangle $  & $\left| \Psi _{T_{e}}\right\rangle $
 & $\left| \Psi _{T_{u}}\right\rangle $ & $\left| \Psi
_{T_{d}}\right\rangle $ \\
\hline $\left\langle n_{3\uparrow }\right\rangle$ & $\frac{1}{2}$
& $\frac{1}{2}(1-4RT\Theta)$ & $1-R\Theta$
 & $R\Theta$
\\
\hline $\left\langle n_{3\downarrow }\right\rangle$ &
$\frac{1}{2}$ & $\frac{1}{2}(1-4RT\Theta)$ & $R\Theta$ &
$1-R\Theta$
\\
\hline $\left\langle n_{4\uparrow }\right\rangle$ &  $\frac{1}{2}$
& $\frac{1}{2}(1-4RT\Theta)$ & $1-T\Theta$ & $T\Theta$
\\
\hline $\left\langle n_{4\downarrow }\right\rangle$ &
$\frac{1}{2}$ & $\frac{1}{2}(1-4RT\Theta)$ & $T\Theta$ &
$1-T\Theta$
\\
\hline $\left\langle n_{3\uparrow }n_{3\downarrow }\right\rangle$
& $2RT\cos^{2}2\gamma_{-}$ &
$2RT\cos^{2}\theta\sin^{2}2\gamma_{-}$ & $RT\Theta$ & $RT\Theta$
\\
\hline $\left\langle n_{4\uparrow }n_{4\downarrow }\right\rangle$
& $2RT\cos^{2}2\gamma_{-}$ &
$2RT\cos^{2}\theta\sin^{2}2\gamma_{-}$ & $RT\Theta$ & $RT\Theta$
\\
\hline $\left\langle n_{3\uparrow }n_{4\uparrow }\right\rangle$ &
$\frac{1}{2}\Theta$ & $\frac{1}{2}(T-R)^{2}\Theta$ & $1-\Theta$ &
$0$
\\ \hline
$\left\langle n_{3\downarrow }n_{4\downarrow }\right\rangle$ &
$\frac{1}{2}\Theta$ & $\frac{1}{2}(T-R)^{2}\Theta$ & $0$ &
$1-\Theta$
\\
\hline $\left\langle n_{3\uparrow }n_{4\downarrow }\right\rangle$
&
$\frac{1}{2}(T-R)^{2}\cos^{2}2\gamma_{-}+\frac{1}{2}\cos^{2}\theta\sin^{2}2\gamma_{-}$
&
$\frac{1}{2}\cos^{2}2\gamma_{-}+\frac{1}{2}(T-R)^{2}\cos^{2}\theta\sin^{2}2\gamma_{-}$
& $T^{2}\Theta$ & $R^{2}\Theta$
\\ \hline
$\left\langle n_{3\downarrow }n_{4\uparrow }\right\rangle$ &
$\frac{1}{2}(T-R)^{2}\cos^{2}2\gamma_{-}+\frac{1}{2}\cos^{2}\theta\sin^{2}2\gamma_{-}$
&
$\frac{1}{2}\cos^{2}2\gamma_{-}+\frac{1}{2}(T-R)^{2}\cos^{2}\theta\sin^{2}2\gamma_{-}$
& $R^{2}\Theta$ & $T^{2}\Theta$
\\ \hline\hline

\end{tabular}
\end{table*}
\end{center}

\begin{center}
\begin{table*}[htbp]

\caption{The output variances of electrons for input singlet and
triplets ($\Theta=\sin ^{2}\theta \sin ^{2}2\gamma
_{-}$)}\label{tstla2}
\begin{tabular}{|l|c|c|c|c|}  \hline\hline
 & $\left| \Psi _{S}\right\rangle $  & $\left| \Psi _{T_{e}}\right\rangle $
 & $\left| \Psi _{T_{u}}\right\rangle $ & $\left| \Psi
_{T_{d}}\right\rangle $ \\
\hline $\left\langle \Delta n_{3\uparrow }^{2}\right\rangle$ &
$\frac{1}{4}$ & $\frac{1}{4}(1-16R^{2}T^{2}\Theta^{2})$ &
$R\Theta(1-R\Theta)$ & $R\Theta(1-R\Theta)$
\\
\hline $\left\langle \Delta n_{3\downarrow }^{2}\right\rangle$ &
$\frac{1}{4}$ & $\frac{1}{4}(1-16R^{2}T^{2}\Theta^{2})$ &
$R\Theta(1-R\Theta)$ & $R\Theta(1-R\Theta)$
\\
\hline $\left\langle \Delta n_{4\uparrow }^{2}\right\rangle$ &
$\frac{1}{4}$ & $\frac{1}{4}(1-16R^{2}T^{2}\Theta^{2})$ &
$T\Theta(1-T\Theta)$ & $T\Theta(1-T\Theta)$
\\
\hline $\left\langle \Delta n_{4\downarrow }^{2}\right\rangle$ &
$\frac{1}{4}$ & $\frac{1}{4}(1-16R^{2}T^{2}\Theta^{2})$ &
$T\Theta(1-T\Theta)$ & $T\Theta(1-T\Theta)$
\\
\hline $\left\langle \Delta n_{3\uparrow }\Delta n_{3\downarrow
}\right\rangle$ & $2RT\cos^{2}2\gamma_{-}-\frac{1}{4}$ &
$2RT(\sin^{2}2\gamma_{-}-2RT\Theta^{2})-\frac{1}{4}$ &
$-R^{2}\Theta(1-\Theta)$ & $-R^{2}\Theta(1-\Theta)$
\\
\hline $\left\langle \Delta n_{4\uparrow }\Delta n_{4\downarrow
}\right\rangle$ & $2RT\cos^{2}2\gamma_{-}-\frac{1}{4}$ &
$2RT(\sin^{2}2\gamma_{-}-2RT\Theta^{2})-\frac{1}{4}$ &
$-T^{2}\Theta(1-\Theta)$ & $-T^{2}\Theta(1-\Theta)$
\\
\hline $\left\langle \Delta n_{3\uparrow }\Delta n_{4\uparrow
}\right\rangle$ & $ -\frac{1}{4}(1-2\Theta)$ &
$-\frac{1}{4}(1-2\Theta)-4R^{2}T^{2}\Theta^{2}$ & $-RT\Theta^{2}$
& $-RT\Theta^{2}$
\\
\hline $\left\langle \Delta n_{3\downarrow }\Delta n_{4\downarrow
}\right\rangle$ & $ -\frac{1}{4}(1-2\Theta)$ &
$-\frac{1}{4}(1-2\Theta)-4R^{2}T^{2}\Theta^{2}$& $-RT\Theta^{2}$ &
$-RT\Theta^{2}$
\\
\hline $\left\langle \Delta n_{3\uparrow }\Delta n_{4\downarrow
}\right\rangle$ & $\frac{1}{4}(1-2\Theta)-2RT\cos^{2}2\gamma_{-}$&
$\frac{1}{4}(1-2\Theta)+2RT(2\Theta-\sin^{2}2\gamma_{-}-2TR\Theta^{2})$
& $-RT\Theta(1-\Theta)$ & $-RT\Theta(1-\Theta)$
\\
\hline $\left\langle \Delta n_{3\downarrow }\Delta n_{4\uparrow
}\right\rangle$ & $\frac{1}{4}(1-2\Theta)-2RT\cos^{2}2\gamma_{-}$&
$\frac{1}{4}(1-2\Theta)+2RT(2\Theta-\sin^{2}2\gamma_{-}-2TR\Theta^{2})$&
$-RT\Theta(1-\Theta)$ & $-RT\Theta(1-\Theta)$
\\

\hline\hline
\end{tabular}
\end{table*}
\end{center}

\begin{center}
\begin{table*}[htbp]
\caption{The input correlations, average number and variances of
electrons for singlet and triplets}\label{tstla3}
\begin{tabular}{|l|c|c|c|c|}  \hline\hline
 & $\left| \Psi _{S}\right\rangle $  & $\left| \Psi _{T_{e}}\right\rangle $
 & $\left| \Psi _{T_{u}}\right\rangle $ & $\left| \Psi
_{T_{d}}\right\rangle $ \\
\hline $\left\langle n_{1\uparrow}\right\rangle (\left\langle
\Delta n_{1\uparrow }^{2}\right\rangle)$ & $\frac{1}{2}
(\frac{1}{4})$ & $\frac{1}{2} (\frac{1}{4})$ & $1 (0)$
 & $0 (0)$ \\
\hline $\left\langle n_{1\downarrow } \right\rangle (\left\langle
\Delta n_{1\downarrow }^{2}\right\rangle )$ &
$\frac{1}{2}(\frac{1}{4})$ & $\frac{1}{2}(\frac{1}{4})$ &
$0(0)$ & $1(0)$  \\
\hline $\left\langle n_{2\uparrow }\right\rangle  (\left\langle
\Delta n_{2\uparrow }^{2}\right\rangle )$ &
$\frac{1}{2}(\frac{1}{4})$ & $\frac{1}{2}(\frac{1}{4})$ &
$1(0)$ & $0(0)$ \\
\hline $\left\langle n_{2\downarrow }\right\rangle  (\left\langle
\Delta n_{2\downarrow }^{2}\right\rangle )$ &
$\frac{1}{2}(\frac{1}{4})$ & $\frac{1}{2}(\frac{1}{4})$ & $0(0)$ &
$1(0)$ \\
\hline $\left\langle n_{1\uparrow }n_{1\downarrow }\right\rangle
(\left\langle \Delta n_{1\uparrow }\Delta n_{1\downarrow
}\right\rangle )$ & $0(-\frac{1}{4})$ & $0(-\frac{1}{4})$ &
$0(0)$ & $0(0)$ \\
\hline $\left\langle n_{2\uparrow }n_{2\downarrow }\right\rangle
(\left\langle \Delta n_{2\uparrow }\Delta n_{2\downarrow
}\right\rangle )$ & $0(-\frac{1}{4})$ & $0(-\frac{1}{4})$ & $0(0)$
& $0(0)$
 \\
\hline $\left\langle n_{1\uparrow }n_{2\uparrow }\right\rangle
(\left\langle \Delta n_{1\uparrow }\Delta n_{2\uparrow
}\right\rangle )$ & $0(-\frac{1}{4})$ & $0(-\frac{1}{4})$ & $1(0)$
& $0(0)$
\\
\hline $\left\langle n_{1\downarrow }n_{2\downarrow }\right\rangle
 (\left\langle \Delta n_{1\downarrow }\Delta n_{2\downarrow
}\right\rangle )$ & $0(-\frac{1}{4})$ & $0(-\frac{1}{4})$ & $0(0)$
& $1(0)$
\\

\hline $\left\langle n_{1\uparrow }n_{2\downarrow }\right\rangle
(\left\langle \Delta n_{1\uparrow }\Delta n_{2\downarrow
}\right\rangle)$ & $\frac{1}{2}(\frac{1}{4})$ &
$\frac{1}{2}(\frac{1}{4})$ & $0(0)$ & $0(0)$
\\
\hline $\left\langle n_{1\downarrow }n_{2\uparrow }\right\rangle
(\left\langle \Delta n_{1\downarrow }\Delta n_{2\uparrow
}\right\rangle )$ & $\frac{1}{2}(\frac{1}{4})$ &
$\frac{1}{2}(\frac{1}{4})$ & $0(0)$ & $0(0)$
\\
\hline\hline
\end{tabular}
\end{table*}
\end{center}

\section*{Acknowledgments}

This work was supported by the China Natural Science Foundation No.
10274069, 60471052,10225419; the Zhejiang Provincial Natural Foundation
M603193; the National Labratory of Solid State Microstructures:
M031802.

\end{document}